
\input jnl
\line{\hfill UCSBTH-92-39}
\line{\hfill hep-th@xxx/9210017}

\def\nnn{/\kern -0.60em \nabla}
\def\ppl{\partial_+}
\def\pmi{\partial_-}

\title    SUPERSYMMETRY AND POSITIVE ENERGY
          IN CLASSICAL AND QUANTUM TWO-DIMENSIONAL DILATON GRAVITY

\author   {Youngchul Park\footnote{*}{young@denali.physics.ucsb.edu} and
           Andrew Strominger\footnote{\dag}{andy@denali.physics.ucsb.edu}}

\affil \ucsb      

\abstract
\singlespace
An ${N} = {1}$ supersymmetric version of two dimensional dilaton gravity
coupled to matter is considered. It is shown that the linear dilaton vacuum
spontaneously breaks half the supersymmetries, leaving broken a linear
combination of left and right supersymmetries which squares to time
translations. Supersymmetry suggests a spinorial expression for the ADM
energy ${M}$, as found by Witten in four-dimensional general relativity.
Using this expression it is proven that ${M}$ is non-negative for smooth
initial data asymptotic (in both directions) to the linear dilaton vacuum,
provided that the (not necessarily supersymmetric) matter stress tensor
obeys the dominant energy condition. A {\it quantum} positive energy theorem
is also proven for the semiclassical large-${N}$ equations, despite the
indefiniteness of the quantum stress tensor. For black hole spacetimes, it is
shown that $M$ is bounded from below by $e^{- 2 \phi_H}$, where $\phi_H$ is
the value of the dilaton at the apparent horizon, provided only that the
stress tensor is positive outside the apparent horizon. This is the
two-dimensional analogue of an unproven conjecture due to Penrose. Finally,
supersymmetry is used to prove positive energy theorems for a large class of
generalizations of dilaton gravity which arise in consideration of the
quantum theory.
\endtitlepage

\centerline{\bf 1. INTRODUCTION}

Two dimensional dilaton gravity has recently been found to be a useful
laboratory for studying quantum gravity in a simplified
context [\cite {cghs, jhas}]. Before coupling to matter,
it is a theory with no local degrees of
freedom: the two gauge degrees
of freedom and two constraints eat up the three components of the metric
together with the dilaton.\footnote{*}{By the same counting
Liouville gravity without a dilaton
has minus one degrees of freedom: a curious fact
which obscures analogies to four-dimensional gravity.}  Thus one can
study the interesting global aspects of gravity in isolation from the
complicated dynamics of propagating gravitons.

In adopting two-dimensional dilaton gravity as a model for
four-dimensional gravity, it is important to know what features the two
theories have in common.
For example, is there a positive energy theorem for the two-dimensional
case? In the present paper we address this question following Witten's
supersymmetric proof of the positive energy theorem
in four dimensions [\cite{ewit, ghas, jnes}].
The basic idea is simple: in a supersymmetric theory, ${H} = {Q^2}$
and so must be positive. In section 2 we review
the supersymmetric version of dilaton gravity,
show that the linear dilaton vacuum is the unique supersymmetric solution
and that this vacuum preserves
a non-chiral combination of the supersymmetries.
In section 3 we prove that all smooth
initial data with non-negative stress tensors on a spacelike slice asymptotic
to the linear dilaton vacuum (in both directions) have positive energy.
In section 4 positivity is proven for a scalar matter sector governed
by the action
${S_Z} = {\int} \,({-} {1 \over 2} ({\nabla}{Z})^2 + {QRZ})$, despite
the fact that the associated stress tensor is indefinite.
The Bondi mass is also shown to be positive, after noting a correction
term linear in the $Z$ field. This result is
then used in section 5 to prove a quantum positive energy theorem
for the large-${N}$ equations of [\cite{cghs}].
While it is generally believed (or hoped) that positivity
of the total energy remains valid at the quantum level, this is
apparently the first example for which a theorem has been established.
In section 6, we consider initial data on spacelike slices bounded
by an apparent horizon on one end and asymptotic to the linear dilaton
vacuum on the other. This corresponds to a black hole. A simple proof
is given that in this case the mass is bounded by ${e^{-2\phi_H}}$, where
${\phi_H}$ is the value of the dilaton at the horizon. This establishes
the two-dimensional analogue of Penrose's conjecture that the mass of a
three-dimensional initial data set is bounded by the square root
of the area of the horizons, whose validity is related to
cosmic censorship [\cite{rpen}]. Finally
in section 7 the most general supersymmetric power-counting renormalizable
theory of dilaton gravity involving three arbitrary functions of the
dilaton coupled to matter is considered. Supersymmetry is used to prove a
positive energy theorem for a large subset of these theories.
In conclusion, positivity of the energy is a robust feature of dilaton
gravity. This increases
our confidence that two-dimensional dilaton gravity is a good toy model for
four-dimensional gravity.

The possibility of further applications to the quantum theory is a
key motivation for our investigations.
Recently it has become clear that better control
over higher-order quantum corrections
is essential for understanding the problem
of two-dimensional
black hole formation/evaporation. Typically (extended) supersymmetry
has been very useful in this regard. For example our result that the linear
dilaton vacuum is supersymmetric strongly suggests that it is an
exact solution of the full quantum theory.

\centerline{\bf 2. SUPERSYMMETRIC DILATON GRAVITY}

The $N=1$ supersymmetric extension of dilaton gravity can be
worked out using the superfields found by Howe [\cite{howe}],
whose notation we follow, except for the sign of ${R}$.
The supersymmetric version
of a closely related theory has been described by Chamseddine [\cite{cham}].
Supersymmetric dilaton gravity is described by the superspace action
$$
{S}_G = {{i}\over{2\pi}} {\int} {d^2} {x} {d^2} {\theta}
        {E} {e^{-2\Phi}} \biggl[{S} + 2 i D_\alpha \Phi D^\alpha \Phi
          - 4 \lambda \biggr], \eqno(ssg)
$$
\noindent where the superfields are given by [\cite{howe}]
$$\eqalignno{
{\Phi}&={\phi} + {i}{\bar\theta}{\Lambda}
  + {{i}\over{2}}{{\bar\theta}\theta}{F}, \cr
{E} &= {e} \biggl[{1} + {{i}\over{2}}{\bar\theta}
             {{\gamma}^{\mu}} {\chi_\mu}
                + {{\bar\theta}\theta} \biggl({{i}\over{4}} A
                + {{1}\over{8}} {{\epsilon}^{\mu \nu}}
                  {{\bar \chi}_\mu}{\gamma_5}
                       {\chi_\nu}\biggr)\biggr], \cr
{S}&= {A} + {{\bar\theta}}{\Psi}
               + {{i}\over{2}}{{\bar\theta}\theta}{C}, \cr
{C}&= - {R} {-} {{1}\over{2}} {{\bar{\chi}}_{\mu}}
               {{\gamma}^{\mu}}{\Psi} +
            {{i}\over{4}} {{\epsilon}^{\mu\nu}}{{\bar{\chi}}_\mu}
             {\gamma_5}{{\chi}_\nu} {A}
               {-}{{1}\over{2}}{{A}^2}, \cr
{\Psi} &= - {2}{i}{{\epsilon}^{\mu\nu}}{\gamma_5}{{D}_\mu}
               {{\chi}_\nu} - {{i}\over{2}}
               {{\gamma}^\mu}{{\chi}_\mu}{A}, \cr
{{E_\mu}^{a}} &= {{{e}_{\mu}}^{a}} + {i}{\bar \theta}
                  {{\gamma}^{a}}{{\chi}_\mu}
                  + {{i}\over 4} {\bar \theta}{\theta}{{{e}_{\mu}}^{a}}
                       {A}, \cr
{{E_\mu}^{\alpha}} &= {1 \over 2} {{{\chi}_\mu}^{\alpha}}
     + {1\over 2}
          {({\bar \theta} \gamma_5)}^{\alpha} {\omega}_\mu
        - {1 \over 4} {({\bar \theta} {\gamma}_\mu)}^{\alpha} {A}
        - {\bar \theta} \theta
           \biggl({{3 i} \over 16} {{{\chi}_\mu}^{\alpha}} {A}
               + {1 \over 4} {({\gamma}_\mu
                       {\Psi})^\alpha} \biggr), \cr
{{E_\beta}^{a}} &= {i} ({{\bar \theta} {{\gamma}^{a}}})_\beta, \cr
{{E_\beta}^{\alpha}} &= {{\delta_\beta}^{\alpha}}
        \biggl(1 - {i \over 8} {\bar \theta} \theta {A}
\biggr), &({sfd}) \cr}
$$
\noindent and
$$\eqalignno{
{{D}_\mu {\chi}_\nu} &= \partial_\mu {\chi}_\nu
                - {1 \over 2} {{\omega}_\mu}
                     {\gamma_5} {\chi}_\nu, \cr
{{\omega}_\mu} &= - {{e}_{a \mu}}{{\epsilon}^{\nu\rho}}
                {\partial_\nu}{{{e}_\rho}^{a}}
                   + {1 \over 2} {\bar{\chi}_\mu}{\gamma_5}
                    {{\gamma}^{\nu}} {{\chi}_\nu}.&({cnc}) \cr }
$$
$a,b$ are tangent space indices, $\mu, \nu$ are spacetime indices,
$\alpha, \beta$ are spinor indices, ${\epsilon_{01}} = {1}$,
and $D$ is the superderivative.
All spinors are Majorana.
``$\gamma_5$'' is, in a notational
abuse, equal to $\gamma^0 \gamma^1$.  Further details can be found
in [\cite{howe}].

The bosonic part of the action in component form is
$$
S_G = {1 \over {2 \pi}} {\int}{d^2} {x} e {e^{-2\phi}}
             [R  + {4} ({\nabla}{\phi})^2
             + {2 A F} - 4 F^2 + 2 \lambda A - 8 \lambda F]. \eqno(actn)
$$
The equations of motion for the auxilliary fields ${A}$ and ${F}$ are
$$
\eqalignno{
A &= 0,\cr
F &= - \lambda. &({afeq})\cr}
$$
Substituting into (\call {actn}) one finds
$$
{S_G} = {{1}\over{2\pi}} {\int}{d^2}{x}{ee^{-2\phi}} [{R} + {4} ({\nabla}
{\phi})^2 + {4} {\lambda^2}]. \eqno(ctg)
$$
The supersymmetry variations of the fermi fields are
$$\eqalignno{
{\delta}{\Lambda} &= \biggl({\nnn} {\phi} + F \biggr) {\eta}, \cr
{\delta}{\chi_\mu} &= \biggl({2}{D_\mu} + {{1}\over{2}} {\gamma_\mu}{A}
                 \biggr) {\eta}.
&(sur) \cr}
$$
A supersymmetric field configuration is one for which all
supersymmetry variations vanish. Setting the background fermi fields to
zero and the the auxilliary fields to their constant values (\call {afeq}),
this implies
$$\eqalignno{
({\nnn}{\phi} -\lambda){\eta} &= 0, &(sucon) \cr
{2} {D_\mu} {\eta} &= 0.
&(pme) \cr}
$$
The integrability condition for (\call {pme}) is that the
curvature vanishes and the metric is therefore
flat. (\call {sucon}) then implies that $(\nabla \phi)^2 = \lambda^2$.
By an appropriate choice of coordinates the
general solution of (\call {sucon})
can then be written in the form
$$
{\phi} = - {\lambda}{\sigma},~~~~~~~~~~
{g_{\mu\nu}} = {\eta_{\mu\nu}}, \eqno(ldv)
$$
\noindent where ${\sigma}$ is a spatial coordinate
in the ``$1$'' direction. The spinor ${\eta}$
for which the variations vanish obeys (in these coordinates)
$$
({\gamma^{1}} + {1})\,\,{\eta} = {0},~~~~~~~~~~{\eta} = {constant}.\eqno(etas)
$$
\noindent This is true for one combination of the
original two supersymmetries.

The solution (\call {ldv}) is known as the linear dilaton vacuum, and will be
refered to herein simply as the vacuum. It
spontaneously breaks Poincare invariance (down to time translations)
as well as half of the supersymmetries.
Note that neither left nor right moving supersymmetry alone is
unbroken: only a linear combination of the two which squares to time
translations.

\break
\centerline {\bf 3. POSITIVE ENERGY FOR ASYMPTOTICALLY FLAT SPACETIMES}

The complete solution space of pure dilaton gravity (\call {ctg})
consists of the vacuum together with a one parameter family
of black hole solutions [\cite{witt}]. Non-singular initial data
exist for the black holes only if the mass is
positive, so a positive energy theorem is rather trivially
demonstrated.

More generally one would like to know if the energy remains positive when
dilaton gravity is coupled to a general matter theory
with a positive stress tensor.
(Positivity was shown by construction for some special
circumstances in [\cite {cghs}].)
In this section we shall show that this is indeed the case
if the spacetime is asymptotic
to the vacuum (\call {ldv}) in both spatial directions.
In the next section we shall
relax this condition and prove a positive energy theorem
for black hole spacetimes with matter.

Since the vacuum leaves one supersymmetry unbroken, a conserved global
supercharge can be constructed for configurations which are asymptotic
to that vacuum as ${\sigma}{\rightarrow}{\pm}{\infty}$. The standard
Noether procedure leads to
$$\eqalignno{
Q &= i \int d \sigma^\mu
          {\nabla_\mu} \bigl[ {e^{-2\phi}}
         {\bar\eta}{\gamma_5}{\Lambda} \bigr] \cr
   &={i}{e^{-2\phi}}
         {\bar\eta}{\gamma_5}{\Lambda}\Big|^{+\infty}_{-\infty}.
&({scg})\cr}
$$
The integral is over an arbitrary spacelike slice and $d\sigma$
is the line element. $\eta$ is any spinor
asymptotically (in both directions) obeying (\call {etas}) or, in
a general coordinate system,
(\call {sucon}) and (\call {pme}).

Since the square of this supercharge is a time
translation, a Witten-like expression for the ADM mass can be obtained
as a supersymmetry variation of the supercharge. This leads to
$$\eqalignno{
{M} &= \int d {\sigma^\mu} {\nabla_\mu} \bigl[
         2 {e^{-2 \phi}} {\bar \epsilon} {\gamma_5}
            {\delta_\epsilon \Lambda} \bigr] \cr
    &= {2}{e^{{-} 2 \phi}}{\bar\epsilon}{\gamma_5}
        ({\nnn}{\phi} - \lambda)
         {\epsilon}\Big|_{+\infty}, &({mcg})\cr}
$$
\noindent where $\epsilon$ is any {\it commuting} spinor obeying the analog
of the
asymptotic conditions (\call {etas}) and normalized so that
$$
 - \bar\epsilon\gamma_5\epsilon\Big|_{\infty}=1, \eqno(nrm)
$$
\noindent and $\delta_\epsilon$ denotes the supersymmetry variations
(\call{sur}) with $\eta$ replaced by $\epsilon$.
Note that with these asymptotic conditions the boundary term at
$\sigma = -\infty$ in the
expression for $M$ vanishes and has been omitted in (\call {mcg}).

This may be compared with previous expressions for $M$ [\cite {witt}] by
linearizing around the vacuum. Defining
$$\eqalignno{
{\phi} &= {-}{\lambda}{\sigma} + {\delta \phi},\cr
{g_{\mu \nu}} &= \eta_{\mu \nu} + h_{\mu \nu}, &({fvar})\cr}
$$
one finds
$$
M = e^{2 \lambda \sigma} ( 2 \partial_1 \delta \phi + \lambda h_{1 1}) \,
\Big|_{+\infty}\eqno(mprv)
$$
in agreement with previous results.

Using the equation of motion
$$
 {{2 \pi} \over \sqrt{-g}} {{\delta {S_G}} \over
                 {{\delta}{g^{\mu\nu}}}}=
 - {{2 \pi} \over \sqrt{-g}} {{\delta {S_m}} \over
                  {\delta {g^{\mu \nu}}}} \equiv {T_{\mu \nu}},\eqno(strss)
$$
where ${T_{\mu \nu}}$ is a general matter stress tensor, one finds
$$\eqalignno{
M &= \int d {\sigma^\mu} {\nabla_\mu} \bigl[
       2 {e^{-2 \phi}} {\bar \epsilon} {\gamma_5}
         (\nnn \phi - \lambda) \epsilon \bigr] \cr
  &= \int d {\sigma^\mu} \bigl[ {\epsilon_\mu}^\nu
           {\bar \epsilon} {\gamma^\rho} \epsilon T_{\nu \rho}
           + 2 {e^{-2 \phi}}
              {{({\delta_\epsilon \bar\Lambda})}} \gamma_5
                \delta_\epsilon {\widetilde \chi}_\mu \bigr],&({mmcg})\cr}
$$\noindent where
$$
\delta_\epsilon {\widetilde \chi}_\mu = \delta_\epsilon {\chi_\mu}
                  - \gamma_\mu \delta_\epsilon \Lambda.
$$
\noindent This expression is valid for any choice of $\epsilon$
satisfying the boundary conditions. Positivity can be made manifest by
choosing an epsilon which obeys
$$
\delta_\epsilon {\widetilde \chi}_1 =  (2 D_1 - {\gamma_1} {\nnn \phi}
                     + {\gamma_1} \lambda) {\epsilon} = 0,\eqno(chc)
$$
where ``$1$'' denotes the direction tangent to the spacelike slice.
This is a first order differential equation which may be solved for
$\epsilon$.
The energy is then simply
$$
  M =   \int d {\sigma^\rho} {\epsilon_\rho}^\nu
           {\bar \epsilon} {\gamma^\mu} \epsilon T_{\nu \mu}.\eqno(mfin)
$$
A Fierz identity implies that
$$
 {\bar \epsilon} {\gamma^\rho} \epsilon {\bar \epsilon} {\gamma_\rho} \epsilon
=-( {\bar \epsilon} {\gamma_5} \epsilon)^2, \eqno(frz)
$$
so that $ {\bar \epsilon} {\gamma^\rho} \epsilon$ is a timelike vector.
At infinity it is future directed, and must be so everywhere by continuity.
The expression (\call {mfin}) for $M$ is then manifestly non-negative
as long as ${T_{\mu \nu}}$
satisfies the dominant energy condition.

Thus we have proven that the ADM energy for spacetimes asymptotic to the
vacuum is always non-negative if the matter has non-negative stress tensor,
and vanishes if and only if the matter stress tensor does.
While supersymmetry has been used to motivate expression (\call {mcg}),
the result applies to a much broader class of theories.

\centerline{\bf 4. POSITIVE ENERGY FOR NON-MINIMALLY}
\centerline{\bf COUPLED CONFORMAL SCALARS}

In this section we shall consider dilaton gravity coupled to matter governed
by the action
$$
{S_Z} = {{1}\over{2\pi}} {\int} {d^2}{x} {\sqrt{-g}}\,\,({{-} {1 \over 2}}
({\nabla}{Z})^2 + {QRZ}).\eqno(mtn)
$$
\noindent where ${Q}$ is an arbitrary constant. The associated stress tensor
is
$$
{T_{\mu\nu}^Z} = {1 \over 2} {\nabla_\mu} Z {\nabla_\nu} Z
     - {1 \over 4}
{g_{\mu \nu}} (\nabla Z)^2
   + Q {\epsilon_\mu}^{\alpha} {\epsilon_\nu}^{\beta} {\nabla_\alpha}
{\nabla_\beta} Z. \eqno(nst)
$$
\noindent ${T^Z}$ has no particular positivity properties. The last term,
which dominates for small ${Z}$, changes sign under ${Z}{\rightarrow}
{-}{Z}$. Thus one might not expect a positive energy theorem for the
gravity-${Z}$ system. On the other hand, it is easy to see that
(\call {mtn}) can be supersymmetrized, so the ${H} = {Q^2}$ relation
suggests such a theorem nevertheless exists. In this section we shall
prove that this is indeed the case.

The presence of a second derivative term in ${T^Z}$ implies a correction
to the boundary formula for the mass of the gravity-${Z}$ system:
$$
{M} = [{2}{e^{-2\phi}}{\bar\epsilon}{\gamma_5} ({\nnn}{\phi} - {\lambda})
{\epsilon} - {Q}{\bar\epsilon}{\gamma_5}{\nnn}{Z}{\epsilon}]_{\infty}.
\eqno(mcp)
$$
\noindent For spinors obeying the boundary conditions (\call {etas}),
the extra
term is proportional to the spatial derivative of ${Z}$. If ${M}$ is
evaluated at spatial infinity, and ${Z}$ is asymptotically constant, this
extra term vanishes.

An expression for the Bondi mass is obtained simply by evaluating
(\call {mcp}) at
right future null infinity $({\cal I}_R^+)$. In that case, however, the
extra boundary term is not negligible. Asymptotically, ${Z}$ obeys the free
wave equation and is given by ${Z} = {Z_+}({x^+}) + {Z_-}({x^-})$. Thus in
general if there is outgoing (incoming) ${Z}$ radiation, ${\pmi}
{Z} ({\ppl}{Z})$ will not vanish on ${\cal I}_R^+ ({\cal I}_R^-)$,
and the extra boundary term will be non-zero.

Integrating by parts, the mass formula can be written as in (\call {mmcg}),
$$
{M}= {\int} {d}{\sigma^\mu} [{\epsilon_\mu}\,^{\nu} {\bar\epsilon}
{\gamma^\rho} {\epsilon} {T_{\nu\rho}^Z}
+ 2 {e^{-2\phi}} {\delta_\epsilon}{\bar\Lambda}{\gamma_5}{\delta_\epsilon}
{\widetilde\chi}_\mu
- {Q}{\nabla_\mu} ({\bar\epsilon}{\gamma_5}{\nnn}{Z}{\epsilon})].
\eqno(mmcp)
$$
\noindent The terms involving two derivatives of ${Z}$ cancel
(using ${\bar \epsilon} \gamma_5 \gamma_\mu \epsilon = -
{\epsilon_\mu}^\nu {\bar \epsilon} \gamma_\nu \epsilon$), yielding
$$
{M}= {\int} {d} {\sigma^\mu} [{\epsilon_\mu}\,^{\nu} {\bar\epsilon}
{\gamma^\rho} {\epsilon} {\widehat T}_{\nu\rho}^{Z}
+ 2 {e^{-2\phi}}{\delta_\epsilon}{\bar\Lambda}{\gamma_5} {\delta_\epsilon}
{\widetilde\chi}_\mu
- {Q}{\nabla_\rho}{Z}{\nabla_\mu} ({\bar\epsilon}{\gamma_5}{\gamma^\rho}
{\epsilon})],\eqno(mmp)
$$
\noindent where the reduced stress tensor
$$
{\widehat T}_{\mu\nu}^Z = {{1}\over{2}} {\nabla_\mu}{Z}{\nabla_\nu}{Z}
- {{1}\over{4}} {g_{\mu\nu}} ({\nabla}{Z})^2, \eqno(trd)
$$
\noindent obeys the dominant energy condition.

The gauge choice (\call {chc}) for ${\epsilon}$ is not useful for
proving positivity of the gravity-${Z}$ system. We choose instead the
modified condition
$$
{\delta_{\epsilon}}{\widetilde\chi}_1
     = {1 \over 2} Q e^{2\phi} \gamma_1 \nnn Z.
\eqno(eee)
$$
\noindent One then finds after some algebra
$$
{M} = {\int} {d} {\sigma^\rho} {\epsilon_\rho}^\mu
      (1 - 2 {Q^2} {e^{2\phi}}) {\bar \epsilon}
{\gamma^\nu} \epsilon {\widehat T}_{\mu\nu}.\eqno(mtr)
$$
\noindent Since ${\widehat T}_{\mu\rho}$ obeys the dominant energy condition,
this expression is non-negative as long as ${Z}$ has support only in the
region where
$$
2 {Q^2} {e^{2\phi}} < 1. \eqno(qqq)
$$
\noindent This last restriction comes as no surprise to those familiar with
the relation between dilaton gravity-${Z}$ system and the large-${N}$
quantum equations, to which we now turn.

\centerline{\bf 5. A QUANTUM POSITIVE ENERGY THEOREM}

The classical gravity-${Z}$ dynamics of the previous section is closely
related to the large-${N}$ quantum dynamics of dilaton gravity
minimally coupled to ${N}$ scalars. To see this, note that the ${Z}$
equation of motion,
$$
- {\square}{Z} = {QR},\eqno(2com)
$$
\noindent can be substituted into the trace of Einstein equations to
yield
$$
2 {e^{-2\phi}} ( - {\square}{\phi} + 2 (\nabla \phi)^2 - 2 {\lambda^2})
      = {Q^2}{R}.\eqno(tcom)
$$
\noindent This is identical to the large-${N}$ quantum trace equation
of [\cite {cghs}] for
$${Q^2} = {N \over {24}}. \eqno(qlgn)
$$
\noindent The dilaton equation is
unaffected by matter, so it is also identical for the two cases.

In addition to these two equations there are the constraint equations
which are most easily expressed in conformal gauge
$$\eqalignno{
{g_{++}}&= {g_{--}} = {0}, \cr
{g_{+-}}&= {-} {{1}\over{2}} {e^{2\rho}}, &({cgg})\cr}
$$
\noindent where ${x^{\pm}} = {x^0}{\pm}{x^1}$. The ${++}$ constraint
equations for the $({Z}, {\rho}, {\phi})$ system is
$$
 0 = {e^{-2 \phi}}(4 \ppl \rho \ppl \phi - 2 \ppl^2 \phi)
         + {T_{++}^Z}, \eqno(2cn)
$$
\noindent where
$$
{T_{++}^Z} = {{1}\over{2}} ({\ppl}{Z})^2
                 + {Q}({\ppl^2} Z - 2 \ppl \rho \ppl Z). \eqno(+++)
$$
\noindent The large-${N}$ constraint equation is
$$
 0 = {e^{-2\phi}} (4 \ppl \rho \ppl \phi - 2 \ppl^2 \phi)
                 + {T_{++}^{Q}} + {T_{++}^M}, \eqno(ncn)
$$
\noindent where ${T^M}$ is the classical matter stress tensor and
$$
{T^Q_{++}} = {-} {{N}\over{12}} (({{\ppl} \rho})^2 -
{\ppl^2}{\rho}) + {t_+} ({x^+}).\eqno(tqm)
$$
\noindent ${t_+}$ is an arbitrary function of ${x^+}$ which is fixed
by boundary conditions. A similar equation holds for ${T_{--}}$. If
$({\rho},{\phi})$ satisfy the dilaton and large-${N}$ trace
equations, it is always possible to find a ${t_{\pm}}$ such
that the ${T_{\pm\pm}}$ constraint equations hold. Since the
dilaton and trace equations are identical (after using the ${Z}$ equations
of motion) for the quantum large-${N}$ and classical gravity-${Z}$ systems,
it follows that every $({\rho}, {\phi}, {Z})$ which satisfy the
classical gravity-${Z}$ equations provide a $({\rho},{\phi})$ which
solve the quantum large-${N}$ equations.

The converse is not always true: given a solution $({\rho}, {\phi})$ of the
quantum equations, it is not always possible to reconstruct a solution
$({Z}, {\rho}, {\phi})$ of the classical equations. Attempts to do so may
run into singularities in the ${Z}$ field. As an example, suppose one
has a solution of the quantum equation such that on the
null slice ${x^-} = {x_0^-}$, $\rho =0$ and
$$
{T_{++}^Q} ({x_0^-}, {x^+}) + {T^M_{++}} ({x_0^-}, {x^+})
= {a} {\delta} ({x^+}). \eqno(tst)
$$
\noindent We wish to find an asymptotically constant function
$\ppl Z  ({x_0^-}, {x^+})$
such that
$$
{{1}\over{2}} ({\ppl}{Z} ({x_0^-}, {x^+}))^2 - {Q}{\ppl^2}
{Z} ({x_0^-}, {x^+}) = {a}{\delta} ({x^+}). \eqno(zad)
$$
\noindent The general solution is
$$
\ppl Z = - 2 Q \biggl( \theta (- x^+ ) { 1 \over
{x^+ + \alpha}} + \theta (x^+) { 1 \over { x^+ + \beta }} \biggr),
\eqno(bbb)
$$
\noindent where
${1 \over \beta} - {1 \over \alpha} = {a \over {2 Q^2}}$.
This is non-singular only if $\alpha < 0$ and
$\beta > 0$, which is possible only if $a > 0$, corresponding to
a positive stress tensor in (\call {tst}).

This is of course expected: one cannot hope to prove positivity
for every solution of the large-${N}$ equations with
unrestricted ${t_\pm}$, since the ${t_\pm}$ can be
chosen to correspond to negative energy matter.

In physical interesting situations the ${t}{\pm}$ are constrained. For
example solutions of the large-${N}$ equations have been studied which
correspond to black hole formation and evaporation. Our results may
be used to show that in these examples
the Bondi mass is always positive, as follows.

A black hole can be formed by specifying that the initial data on
${\cal I}_L^-$ correspond to the vacuum while on ${\cal I}_R^-$
one has some general incoming radiation pulse
$$
{T_{++}^M} = {T_{++}^Q} + {T_{++}^M} > {0}, \eqno(tmt)
$$
\noindent which we take to have compact support. (A shock wave
corresponds to ${T_{++}^M} = {a}{\delta} ({x^+} - {x_0^+})$.) If
${T_{++}^M} + {T_{++}^Q}$ is positive, one can always
construct an asymptotically constant $\ppl Z$ such that
$$
{T_{++}^Z} = {T_{++}^M} + {T_{++}^Q}\eqno(zmq)
$$
\noindent on ${\cal I}_R^-$. The Bondi mass ${M_B}$ on
${\cal I}_R^-$ (i.e., the ADM mass) is thus positive by the
theorem of the preceeding section. The Bondi mass at finite values of
${x^-}$ may then be found from (\call {mtr}), integrating over a null
slice of constant ${x^-}$. It will remain positive as long as
${\ppl}{Z}$ is non-singular on the slice, and has support only
in the region ${e^{-2\phi}}>{{N}\over{12}}$.

When can ${\ppl Z}$ become singular? From the ${Z}$ equation of motion
$$
{\ppl}{\pmi}{Z}
       = 2 Q {\ppl}{\pmi}{\rho}, \eqno(aaa)
$$
\noindent it is evident that if the scalar curvature diverges,
${\ppl Z}$ does
as well. In fact this condition is necessary as well as sufficient.
If ${\ppl Z}$ is initially finite on ${\cal I}_R^-$ and $\ppl Z$
eventually diverges, there must be a first value of ${x^-}$ at which
it does so. Thus ${\ppl}{Z}$ goes from a finite to infinite value in
a finite
interval, so
${\ppl}{\pmi}{Z}$ must diverge, together with the scalar curvature.

We thus conclude that as long as the scalar curvature is finite, and the
${Z}$ pulse has not crossed the line
${e^{-2\phi}} = {{N}\over{12}}$,
${M_B}$ is non-negative.

It has been shown [\cite{bddo, rst}] that
when the leading edge of the pulse first intersects the line
${e^{-2\phi}} = {{N}\over{12}}$, a curvature singularity appears
which then continues to the right along a spacelike trajectory. Thus
every null slice which is prior to the singularity necessarily has ${\ppl}
{Z} = {0}$ in the region ${e^{-2\phi}}<{{N}\over{12}}$, and positivity
therefore follows from finiteness of the scalar curvature alone.

If the sequence of null slices first encounters the curvature
singularity at a finite value of ${x^+}$ (i.e., not on ${\cal I}_R^+$)
it is by definition a naked singularity and the end of the black hole.
Our result then implies that the black hole mass cannot become negative
before it disappears. The behavior of the mass after the naked singularity
will depend on the boundary condition imposed there.

On the other hand it may be the case [\cite{shjs}]
that the null slices meet the singularity
at $ {\cal I}_R^+ $. This is then the end of the spacetime, and our
results imply that the Bondi mass is everywhere positive.

These statements may seem at odds with
references [\cite{qtdg, deal, bica, ept}] in which it
was stated, in modified versions of the large-${N}$ equations, that the Bondi
mass in fact does get negative before the black hole disappears. One
possibility is that an analogous theorem does not exist for the modified
equations. In fact we believe, though we have not worked out the details,
that the methods of section 7 can be used to prove ${M_B}>{0}$
for the modified equations.

More likely we believe the discrepancy lies in the differing definitions of
${M_B}$. Previous work did not include the crucial
$Q \pmi Z$ term,\footnote{*}{Note that if $\rho$ and $Z$ both vanish
on ${\cal I}_L^-$ --- which is true in some gauge for most cases of
interest --- the quantum mass correction is by virtue of (\call {aaa})
proportional to $\pmi \rho$.}
which may be regarded as a quantum correction to the classical
mass formula.

We do not know which is ``the'' correct mass formula. However we note that
supersymmetry suggests that the energy should be positive even at the
quantum level, as we have indeed found to be the case with our modified
${M_B}$.

\centerline{\bf 6. AN ENERGY BOUND FOR BLACK HOLE SPACETIMES}

In four dimensional general relativity there is a conjecture due to
Penrose that the mass ${M_i}$ of an initial data slice containing apparent
horizons is bounded by the square root of the total area ${A_i}$ of
the apparent horizons [\cite {rpen}].
The motivation behind this conjecture is that
eventually the system is expected to settle down to Schwarzchild with
a larger horizon area ${A_f}$ (by the area theorem, which
assumes cosmic censorship) and less energy (having lost some in
outgoing radiation). The final mass ${M_f}$ is then proportional to
the square root
of the final area so that
$$
{M_i} \geq {M_f} = {\sqrt{{A_f}\over{16\pi}}} \geq
            {\sqrt{{A_i}\over{16\pi}}},
$$
\noindent implying Penrose's conjecture,
$M \geq {\sqrt{ A \over {16 \pi}}}$.

This conjecture has been proven only in special cases
in four dimensions [\cite {jawa}].
In this section an analogous inequality will be proven for two-dimensional
dilaton gravity. The expected inequality can be phrased by noting that the
value of ${\lambda}{e^{-2\phi}}$ at the horizon plays the role analogous
to that ${\sqrt{{A}\over{16\pi}}}$
in four dimensions. (Indeed, when two-dimensions
dilaton gravity
is derived by dimensional reduction of spherically symmetric four-dimensional
black holes [\cite{cghs,bddo,dxbh}],
${e^{-4\phi}}$ is proportional to the
area of the two spheres at constant radius.) In the absence of matter,
the black hole mass is exactly ${\lambda}{e^{-2\phi}}\big|_{\rm Horizon}$.
Adding positive energy matter outside the black hole should only
increase the total mass, so one expects a bound
$$
{M}\,{\geq}\,{\lambda}{e^{-2\phi}}\big|_{\rm Horizon}.\eqno(bnd)
$$
Blackhole spacetimes in dilaton gravity are asymptotic to the vacuum only
as ${\sigma}{\rightarrow}{+}{\infty}$, and therefore are not covered
by the analysis of section 3. In this section
we shall prove that such spacetimes satisfy
the energy bound
(\call {bnd}) provided only
that the matter stress tensor is positive outside the
outermost apparent horizon.

An apparent horizon is a line along which the
gradient of the dilaton field becomes null [\cite{SuTh,jhas}].
The existence of an apparent
horizon can be determined from the initial data on a spacelike slice. This
is in contrast to an event horizon, whose location can be determined
only when the entire spacetime is known.

In this section, we will use an alternative formula for the mass
$$
{\widetilde M} = e^{-2 \phi} \bigl(\lambda
                - {1 \over \lambda} (\nabla \phi)^2 \bigr), \eqno(mfld)
$$
\noindent which is similar to a formula employed by Susskind and
Thorlacious [\cite{SuTh}]. It is easy to check that (\call {mfld}) agrees
asymptotically with (\call {mcg}), though the two are not the same
at finite points. In terms of ${\widetilde M}$, the metric
equation of motion (\call {strss}) can be written
$$
2 e^{-2 \phi} {\epsilon_\mu}^\rho {\epsilon_\nu}^\sigma
      \nabla_\rho \nabla_\sigma \phi
          = 2 \lambda g_{\mu \nu} {\widetilde M} + T_{\mu \nu}. \eqno(rum)
$$
\noindent It is then easy to show using (\call {rum}) that
$$
d {\widetilde M} = (d \sigma^\rho {\epsilon_\rho}^\nu)
               \biggl({1 \over \lambda}
                  \epsilon^{\mu \sigma} \nabla_\sigma \phi\biggr)
                     T_{\mu \nu}. \eqno(dlt)
$$
\noindent Since $\epsilon^{\mu \sigma} \nabla_\sigma \phi$ is
timelike and $T_{\mu \nu}$ obeys the dominant
energy condition, this implies that ${\widetilde M}$ increases away
from the horizon. One therefore concludes that ${\widetilde M} \big|_H$
is less than or equal to the ADM mass ${\widetilde M} \big|_{+ \infty}$.
On the other hand, since $(\nabla \phi)^2$ vanishes at a horizon
$$
{\widetilde M} \big|_H = \lambda e^{-2 \phi} \big|_H . \eqno(mte)
$$\noindent
So we have established the desired inequality (\call {bnd}).
Actually, this method easily establishes
positivity of energy for the case of
two asymptotically flat directions
as well, but the spinorial proof was used in
section 3 because it elucidates the connection with supersymmetry, and
naturally generalizes to the models of the sections 4 and 7.

The theorems of this section strengthen the analogy of two-dimensional dilaton
gravity with four-dimensional general relativity. It would certainly be of
interest to formulate and prove (or disprove) cosmic censorship in
this theory.

\centerline{\bf 7. GENERALIZED DILATON GRAVITY}

The action (\call {actn}) is of course a very special form for the
two-dimensional dynamics of a scalar field coupled to gravity. In
many contexts it is of interest to consider a more general form for the
dynamics. At the very least, quantum corrections will generate corrections
to (\call {actn}) as a power series in ${e^{2\phi}}$. Indeed, these
corrections have been found to play a crucial role in blackhole
dynamics [\cite{ruts,deal,bica,qtdg, ept}]. In this section we consider
the most general, power-counting renormalizable, supersymmetric action:
\footnote{*}{Field redefinitions can be used to locally eliminate two of the
three free functions in $S_G$. However since global considerations may be
important we will not do this.}
$$
{S_G} = {i \over {2 \pi}} \int d^2 x d^2 \theta E
         \bigl[J (\Phi) S
          + i K (\Phi) D_\alpha \Phi D^\alpha \Phi
          + L (\Phi) \bigr]. \eqno(gactn)
$$
\noindent Presumably many choices of the functions $J, K$ and $L$ lead to
sick theories with unphysical behavior.
It is a difficult (but interesting) problem to characterize the
dynamics of the general theory.  In this section we will analyze two
key properties of these theories: the existence of a supersymmetric
ground state and positivity of the energy.

We first record the bosonic part of the action following from (\call {gactn}).
The auxilliary field equations of motion are
$$\eqalignno{
{F} &= - {L \over {2 J^{\prime}}}, \cr
{A} &= - {L^{\prime} \over {J^\prime}} + {{2 K L} \over {{J^\prime}^2}},
&({gafeq})\cr}
$$
\noindent where ${J^{\prime}}$ $({L^{\prime}})$ is the derivative
of ${J}$ $(L)$
with respect to ${\phi}$. This leads to the bosonic action
$$
{S_G} = {1 \over {2 \pi}} \,{\int} \,{d^2}{x}{e}
        \biggl[J R + 2 K (\nabla \phi)^2
           + {{L L^\prime} \over {2 J^\prime}}
           - {{K L^2} \over {2 {J^\prime}^2}}
\biggr]. \eqno(bgact)
$$
\noindent The supersymmetry transformation laws are obtained by
substituting (\call {gafeq}) into (\call {sur}).
A supersymmetric vacuum is one
for which there is a spinor ${\epsilon}$ such that
$$
{\delta_\epsilon} {\Lambda} = (\nnn \phi - {L \over{2 J^\prime}}) \epsilon
= 0, \eqno(lgsur)
$$
$$
{\delta_\epsilon}{\chi_\mu} =
        \biggl({2D_\mu} - {\gamma_\mu} \bigl({L^\prime \over {2 J^\prime}}
             - {{K L} \over {{J^\prime}^2}}\bigr) \biggr) \epsilon = 0.
\eqno(cgsur)
$$
\noindent These equations imply that
$$
{k^{\mu}} = {\bar\epsilon}{\gamma^\mu}{\epsilon}\eqno(kdef)
$$
\noindent is a Killing vector:
$$
{\nabla_{(\nu}}{k_{\mu)}}={0},\eqno(kcd)
$$
\noindent which also generates a symmetry of ${\phi}$
$$
{k^{\mu}}{\nabla_\mu}{\phi}={0},\eqno(kccd)
$$
\noindent and is timelike:
$$
{k^2}<{0}.\eqno(kcccd)
$$
\noindent Multiplying in (\call {lgsur}) by ${{{\bar\epsilon}{\gamma_5}}
\over{\sqrt{-k^2}}}$ one finds
$$
{d \over {d \sigma}} \phi = - {L \over {2 J^\prime}}, \eqno(zzx)
$$
\noindent where
$$
{\sigma}({x}) = {\int^x_{-\infty}}\,{dx^{\mu}}
                   \,\,{{{\epsilon_{\mu\nu}}{k^\nu}}
                   \over{\sqrt{-k^2}}}\eqno(sgm)
$$
\noindent is a spatial coordinate. ${\phi}({\sigma})$ is then the solution of
$$
{\sigma} = - \int {{{2 J^\prime}(\phi)} \over {L(\phi)}} d \phi.
\eqno(phf)
$$
\noindent The curvature is determined from the integrability condition for
(\call {cgsur}). Given (\call {lgsur}) one finds
$$
R = - {1 \over 2} (A^2 + 2 A^\prime F), \eqno(crv)
$$\noindent where A and F are given in (\call {gafeq}).

(\call {phf}) will have a solution for every sigma as long as $L / {J^\prime}$
is everywhere finite. The reason for the absence of a non-singular vacuum
if ${J^{\prime}}$ has zeros is evident from (\call {gactn}): the kinetic
part of the action degenerates at zeros of ${J^{\prime}}$. If the
right hand side of (\call {crv}) is everywhere finite, then it
can be solved to determine the geometry. Thus a
supersymmetric vacuum exists in a wide variety of cases.

Following the steps of section 3, an expression for the mass at ${\sigma}
= {+\infty}$ can for spacetimes asymptotic to the vacuum be found as
$$
M (+ \infty) = - {\bar\epsilon} {\gamma_5}
               ({J^\prime} \nnn \phi
               - {L \over 2}) \epsilon, \eqno(gmdef)
$$
\noindent where ${\epsilon}$ is asymptotically a solution of (\call {lgsur})
and (\call {cgsur}) normalized according to (\call {nrm}).
Integrating by parts as in section 3 one
finds
$$
{M}({+\infty}) = {\int} {d}{\sigma^\mu} ( {\epsilon_\mu}^\nu
          {\bar \epsilon} {\gamma^\rho} {\epsilon} T_{\nu \rho}
         - {J^\prime} {\delta_\epsilon}
           {\bar \Lambda} \gamma_5 {\delta_\epsilon}
           {\widetilde \chi}_\mu) + {M} (-\infty), \eqno(zyt)
$$
\noindent where
$$
{\delta_\epsilon} {\widetilde \chi}_\mu = {\delta_\epsilon} {\chi_\mu}
           + \gamma_\mu {K \over {J^\prime}} {\delta_\epsilon} \Lambda.
$$
\noindent Choosing ${\epsilon}$ so that
${\delta_\epsilon} {\widetilde \chi}_1 = 0$
one has
$$
M (+ \infty) = \int d {\sigma^\rho} {\epsilon_\rho}^\nu {\bar \epsilon}
{\gamma^\mu} \epsilon T_{\nu \mu} + M (- \infty).\eqno(npn)
$$
\noindent This is manifestly non-negative if ${M} ({-\infty})$ is.

A sufficient condition for ${M} ({-\infty})$ to vanish is that ${J^{\prime}}
(\phi (-\infty))$ and $L( \phi( -\infty))$ vanish.
This is the case for the model of section 3,
and is true for a wide range of choices of ${J}$ and ${L}$. Alternately, if
$\phi$ reaches a zero of ${L}$ at ${\phi_0}$, it follows from
(\call {zzx}) that it will not cross ${\phi_0}$, and ${J^{\prime}}$ and
${L}$ will asymptotically approach ${J^{\prime}}({\phi_0})$ and ${L}
({\phi_0}) = {0}$. ${M}({-\infty})$ will then vanish, and ${M}({+\infty})$
will be non-negative. We do not know the necessary conditions
for non-negativity of ${M} ({+\infty})$.

We expect that similar results
can be derived for generalized classical equations by adding the $Z$
field as in section 5. This would be relevant to some of
the semiclassical models studied
in [\cite{deal, bica, qtdg, ept}].

\centerline{\bf ACKNOWLEDGEMENTS}

We are grateful to Gary Horowitz for useful discussions.
This work was supported in part by DOE-91ER40618.

\references

\refis{howe} P. S. Howe, ``Super Weyl Transformations in Two Dimensions'',
                J. Phys. A: Math. Gen. {\bf 12}, 393 (1979).

\refis{cham} A. H. Chamseddine, ``Super Strings in Arbitrary Dimensions'',
                Phys. Lett. {\bf 258B}, 97 (1991).

\refis{witt} E. Witten, `` String Theory and Black Holes'',
                Phys. Rev. {\bf D44}, 314 (1991).

\refis{cghs} C. G. Callan, S. B. Giddings, J. A. Harvey
                and A. Strominger, ``Evanescent Black Holes'',
                Phys. Rev. {\bf D45}, R1005 (1992).

\refis{jawa} P. S. Jang and R. M. Wald, `` The Positive Energy Conjecture
                and the Cosmic Censor Hypothesis'',
                J. Math. Phys. {\bf 18}, 41 (1977).

\refis{SuTh} L. Susskind and L. Thorlacius, ``Hawking Radiation and
                Back-reaction'', Stanford preprint SU-ITP-92-12 (1992).

\refis{bddo} T. Banks, A. Dabholkar, M. R. Douglas and M. O'Loughlin,
                ``Are Horned Particles the Climax of Hawking Evaporation?'',
                Phys. Rev. {\bf D45}, 3607 (1992).

\refis{bica} A. Bilal and C. G. Callan, ``Liouville Models of Black Hole
                Evaporation'', Princeton preprint PUPT-1320 (1992),
                hep-th@xxx/9205089.

\refis{ruts} J. G. Russo and A. A. Tseytlin, ``Scalar-Tensor Quantum
                Gravity in Two Dimensions'', Stanford/Cambridge preprint
                SU-ITP-92-2 = DAMTP-1-1992.

\refis{rpen} R. Penrose, Ann. N. Y. Acad. Sci. {\bf 224}, 125 (1973).

\refis{jhas} For recent reviews see
                J. A. Harvey and A. Strominger, ``Quantum Aspects of
                Black Holes'' preprint EFI-92-41, hep-th@xxx/
                9209055 , to appear in the
                proceedings of the 1992 TASI Summer School in Boulder,
                Colorado, and S. B. Giddings, ``Toy Models for
                Black Hole Evaporation'' preprint UCSBTH-92-36,
                hep-th@xxx/9209113, to appear in the proceedings of the
                International Workshop of Theoretical Physics, 6th Session,
                June 1992, Erice, Italy.

\refis{dxbh} S. B. Giddings and A. Strominger, ``Dynamics of Extremal
                Black Holes'', Phys. Rev. {\bf D46}, 627 (1992).

\refis{deal} S. P. deAlwis, ``Quantization of a Theory of 2D Dilaton Gravity'',
                Boulder preprint COLO-HEP-280.

\refis{qtdg} S. B. Giddings and A. Strominger, ``Quantum Theories of
                Dilaton Gravity'', UCSB-TH-92-28, hepth@xxx/9207034.

\refis{ewit} E. Witten, ``A New Proof of the Positive Energy Theorem'',
                Commun. Math. Phys. {\bf 80}, 381 (1981).

\refis{ghas} G. T. Horowitz and A. Strominger, ``Witten's Expression
                for Gravitational Energy'', Phys. Rev. {\bf D27},
                2793, (1982).

\refis{jnes} J. M. Nester, ``A New Gravitational Energy Expression
                with a Simple Positivity Proof'', Phys. Lett. {\bf 83A},
                241, (1981).

\refis{rst}  J. G. Russo, L. Susskind and L. Thorlacius,
                ``Black Hole Evaporation in 1+1 Dimensions'',
                Stanford preprint SU-ITP-92-4 (1992).

\refis{shjs} S. W. Hawking and J.M. Stewart, ``Naked and Thunderbolt
                Singularities in Black Hole Evaporation'',
                hep-th@xxx/9207105.

\refis{ept} J. G. Russo, L. Susskind and L. Thorlacius, ``The Endpoint
                of Hawking Radiation'', Stanford preprint
                SU-ITP-92-17 (1992).

\endreferences

\endit